\input phyzzx
\input tables

\catcode`@=11 % This allows us to modify PLAIN macros.
\def\space@ver#1{\let\@sf=\empty \ifmmode #1\else \ifhmode
   \edef\@sf{\spacefactor=\the\spacefactor}\unskip${}#1$\relax\fi\fi}
\def\attach#1{\space@ver{\strut^{\mkern 2mu #1} }\@sf\ }
\newtoks\foottokens
\newbox\leftpage \newdimen\fullhsize \newdimen\hstitle
\newdimen\hsbody
\newif\ifreduce  \reducefalse
\def\almostshipout#1{\if L\lr \count2=1
      \global\setbox\leftpage=#1 \global\let\lr=R
  \else \count2=2
    \shipout\vbox{\special{dvitops: landscape}
      \hbox to\fullhsize{\box\leftpage\hfil#1}} \global\let\lr=L\fi}
\def\smallsize{\relax
\font\eightrm=cmr8 \font\eightbf=cmbx8 \font\eighti=cmmi8
\font\eightsy=cmsy8 \font\eightsl=cmsl8 \font\eightit=cmti8
\font\eightt=cmtt8
\def\eightpoint{\relax
\textfont0=\eightrm  \scriptfont0=\sixrm
\scriptscriptfont0=\sixrm
\def\rm{\fam0 \eightrm \f@ntkey=0}\relax
\textfont1=\eighti  \scriptfont1=\sixi
\scriptscriptfont1=\sixi
\def\oldstyle{\fam1 \eighti \f@ntkey=1}\relax
\textfont2=\eightsy  \scriptfont2=\sixsy
\scriptscriptfont2=\sixsy
\textfont3=\tenex  \scriptfont3=\tenex
\scriptscriptfont3=\tenex
\def\it{\fam\itfam \eightit \f@ntkey=4 }\textfont\itfam=\eightit
\def\sl{\fam\slfam \eightsl \f@ntkey=5 }\textfont\slfam=\eightsl
\def\bf{\fam\bffam \eightbf \f@ntkey=6 }\textfont\bffam=\eightbf
\scriptfont\bffam=\sixbf   \scriptscriptfont\bffam=\sixbf
\def\tt{\fam\ttfam \eightt \f@ntkey=7 }
\def\caps{\fam\cpfam \tencp \f@ntkey=8 }\textfont\cpfam=\tencp
\setbox\strutbox=\hbox{\vrule height 7.35pt depth 3.02pt width\z@}
\samef@nt}
\def\Eightpoint{\eightpoint \relax
  \ifsingl@\subspaces@t2:5;\else\subspaces@t3:5;\fi
  \ifdoubl@ \multiply\baselineskip by 5
            \divide\baselineskip by 4\fi }
\parindent=16.67pt
\itemsize=25pt
\thinmuskip=2.5mu
\medmuskip=3.33mu plus 1.67mu minus 3.33mu
\thickmuskip=4.17mu plus 4.17mu
\def\thinspace{\kern .13889em }
\def\negthinspace{\kern-.13889em }
\def\enspace{\kern.416667em }
\def\enskip{\hskip.416667em\relax}
\def\quad{\hskip.83333em\relax}
\def\qquad{\hskip1.66667em\relax}
\def\crr{\cropen{8.3333pt}}
\foottokens={\Eightpoint\singlespace}
\def\papersize{\SIZE\OFFSET\skip\footins=\bigskipamount}
\def\SIZE{\hsize=11.8truecm\vsize=17.5truecm}
\def\OFFSET{\voffset=-1.3truecm\hoffset=  .14truecm}
\message{STANDARD CERN-PREPRINT FORMAT}
\def\attach##1{\space@ver{\strut^{\mkern 1.6667mu ##1} }\@sf\ }
\def\PH@SR@V{\doubl@true\baselineskip=20.08pt plus .1667pt minus
.0833pt
             \parskip = 2.5pt plus 1.6667pt minus .8333pt }
\def\author##1{\vskip\frontpageskip\titlestyle{\tencp ##1}\nobreak}
\def\address##1{\par\kern 4.16667pt\titlestyle{\tenpoint\it ##1}}
\def\andaddress{\par\kern 4.16667pt \centerline{\sl and} \address}
\def\abstract{\vskip2\frontpageskip\centerline{\tenrm Abstract}
              \vskip\headskip }
\def\cases##1{\left\{\,\vcenter{\Tenpoint\m@th
    \ialign{$####\hfil$&\quad####\hfil\crcr##1\crcr}}\right.}
\def\matrix##1{\,\vcenter{\Tenpoint\m@th
    \ialign{\hfil$####$\hfil&&\quad\hfil$####$\hfil\crcr
      \mathstrut\crcr\noalign{\kern-\baselineskip}
     ##1\crcr\mathstrut\crcr\noalign{\kern-\baselineskip}}}\,}
\Tenpoint
}
\def\Smallsize{\smallsize\reducetrue
\let\lr=L
\hstitle=8truein\hsbody=4.75truein\fullhsize=24.6truecm\hsize=\hsbody
\output={
  \almostshipout{\leftline{\vbox{\makeheadline
  \pagebody\makefootline}}}\advancepageno
     }
\special{dvitops: landscape}
\def\makeheadline{
\iffrontpage\line{\the\headline}
             \else\vskip .0truecm\line{\the\headline}\vskip .5truecm
\fi}
\def\makefootline{\iffrontpage\vskip  0.truecm\line{\the\footline}
               \vskip -.15truecm\line{\the\date\hfil}
              \else\line{\the\footline}\fi}
\paperheadline={
\iffrontpage\hfil
               \else
               \tenrm\hss $-$\ \folio\ $-$\hss\fi    }
\paperstyle}
%
%%%%%%%%%%%%%%%%%%%%%%%%%%%%%%%%%%%%%%%%%%%%%%%%%%%%%%%%%%%%%%%%%%%%%%
%%
%
% Macros for Automatic Reference Numbering
%
%  All references are in a reference file in the following form:
%
%   \def\AAA{\rrr\AAA{This is reference AAA}}
%   \def\BBB{\rrr\BBB{This is reference BBB}}  ...etc
%
%  Then a reference is called as \BBB, and one may call it
%  as often as desired. Before the first reference occurs
%  one should read the reference file by adding \input (fn); the
%  filetype is assumed to be TEX. At the end of the paper an
%  ordered referencelist is produced by \refout.
%
%  Multiple references are called as
%  \multref\AAA{\BBB\CCC\DDD...\ZZZ},
%  which yields e.g. [1-7].
%  Two references may also be called by \doubref\AAA\BBB, which
%  yields e.g. [5,6].
%
%  All PHYZZX routines remain intact.
%
%%%%%%%%%%%%%%%%%%%%%%%%%%%%%%%%%%%%%%%%%%%%%%%%%%%%%%%%%%%%%%%%%%%%
%
\newcount\referencecount     \referencecount=0
\newif\ifreferenceopen       \newwrite\referencewrite
\newtoks\rw@toks
\def\NPrefmark#1{\attach{\scriptscriptstyle [ #1 ] }}
\let\PRrefmark=\attach
\def\refmark#1{\relax\ifPhysRev\PRrefmark{#1}\else\NPrefmark{#1}\fi}
\def\refend{\refmark{\number\referencecount}}
\newcount\lastrefsbegincount \lastrefsbegincount=0
\def\refsend{\refmark{\count255=\referencecount
   \advance\count255 by-\lastrefsbegincount
   \ifcase\count255 \number\referencecount
   \or \number\lastrefsbegincount,\number\referencecount
   \else \number\lastrefsbegincount-\number\referencecount \fi}}
\def\refch@ck{\chardef\rw@write=\referencewrite
   \ifreferenceopen \else \referenceopentrue
   \immediate\openout\referencewrite=referenc.texauxil \fi}
%
% In \obeyendofline, we say `\let^^M=\relax
{\catcode`\^^M=\active % these lines must end with %
  \gdef\obeyendofline{\catcode`\^^M\active \let^^M\ }}%
%
% In \ignoreendofline, we say `\let^^M=\relax
{\catcode`\^^M=\active % these lines must end with %
  \gdef\ignoreendofline{\catcode`\^^M=5}}
{\obeyendofline\gdef\rw@start#1{\def\t@st{#1} \ifx\t@st\blankend%
\endgroup \@sf \relax \else \ifx\t@st\bl@nkend \endgroup \@sf \relax%
\else \rw@begin#1
\backtotext
\fi \fi } }
{\obeyendofline\gdef\rw@begin#1
{\def\n@xt{#1}\rw@toks={#1}\relax%
\rw@next}}
\def\blankend{}
{\obeylines\gdef\bl@nkend{
}}
\newif\iffirstrefline  \firstreflinetrue
\def\rwr@teswitch{\ifx\n@xt\blankend \let\n@xt=\rw@begin %
 \else\iffirstrefline \global\firstreflinefalse%
\immediate\write\rw@write{\noexpand\obeyendofline \the\rw@toks}%
\let\n@xt=\rw@begin%
      \else\ifx\n@xt\rw@@d \def\n@xt{\immediate\write\rw@write{%
        \noexpand\ignoreendofline}\endgroup \@sf}%
             \else \immediate\write\rw@write{\the\rw@toks}%
             \let\n@xt=\rw@begin\fi\fi \fi}
\def\rw@next{\rwr@teswitch\n@xt}
\def\rw@@d{\backtotext} \let\rw@end=\relax
\let\backtotext=\relax

\newdimen\refindent     \refindent=30pt
\def\refitem#1{\par \hangafter=0 \hangindent=\refindent
\Textindent{#1}}
\def\REFNUM#1{\space@ver{}\refch@ck \firstreflinetrue%
 \global\advance\referencecount by 1 \xdef#1{\the\referencecount}}
\def\refnum#1{\space@ver{}\refch@ck \firstreflinetrue%
 \global\advance\referencecount by 1
\xdef#1{\the\referencecount}\refend}

\def\REF#1{\REFNUM#1%
 \immediate\write\referencewrite{%
 \noexpand\refitem{#1.}}%
\begingroup\obeyendofline\rw@start}
\def\ref{\refnum\?%
 \immediate\write\referencewrite{\noexpand\refitem{\?.}}%
\begingroup\obeyendofline\rw@start}
\def\Ref#1{\refnum#1%
 \immediate\write\referencewrite{\noexpand\refitem{#1.}}%
\begingroup\obeyendofline\rw@start}
\def\REFS#1{\REFNUM#1\global\lastrefsbegincount=\referencecount
\immediate\write\referencewrite{\noexpand\refitem{#1.}}%
\begingroup\obeyendofline\rw@start}
\newif\ifmref  %check multi ref
\newif\iffref  %check first ref
\def\xrefsend{\xrefmark{\count255=\referencecount
\advance\count255 by-\lastrefsbegincount
\ifcase\count255 \number\referencecount
\or \number\lastrefsbegincount,\number\referencecount
\else \number\lastrefsbegincount-\number\referencecount \fi}}
\def\xrefsdub{\xrefmark{\count255=\referencecount
\advance\count255 by-\lastrefsbegincount
\ifcase\count255 \number\referencecount
\or \number\lastrefsbegincount,\number\referencecount
\else \number\lastrefsbegincount,\number\referencecount \fi}}
\def\xREFNUM#1{\space@ver{}\refch@ck\firstreflinetrue%
\global\advance\referencecount by 1
\xdef#1{\xrefend}}
\def\xrefend{\xrefmark{\number\referencecount}}
\def\xrefmark#1{[{#1}]}
\def\xRef#1{\xREFNUM#1\immediate\write\referencewrite%
{\noexpand\refitem{#1 }}\begingroup\obeyendofline\rw@start}%
\def\xREFS#1{\xREFNUM#1\global\lastrefsbegincount=\referencecount%
\immediate\write\referencewrite{\noexpand\refitem{#1 }}%
\begingroup\obeyendofline\rw@start}
\def\rrr#1#2{\relax\ifmref{\iffref\xREFS#1{#2}%
\else\xRef#1{#2}\fi}\else\xRef#1{#2}\xrefend\fi}
\def\multref#1#2{\mreftrue\freftrue{#1}%
\freffalse{#2}\mreffalse\xrefsend}
\def\doubref#1#2{\mreftrue\freftrue{#1}%
\freffalse{#2}\mreffalse\xrefsdub}
\referencecount=0
\def\refbreak{\hfil\penalty200\hfilneg}
\def\paperstyle{\papers}
\paperstyle   %  This is the default
%
%%%%%%%%%%%%%%%%%%%%%%%%%%%%%%%%%%%%%%%%%%%%%%%%%%%%%%%%%%%%%%%%%%%%
%
%  Local, ie. site-dependent macros.
%
\def\slacpub{\afterassignment\slacp@b\toks@}
\def\slacp@b{\edef\n@xt{\Pubnum={\the\toks@}}\n@xt}
\let\pubnum=\slacpub
\expandafter\ifx\csname eightrm\endcsname\relax
    \let\eightrm=\ninerm \let\eightbf=\ninebf \fi

\font\seventeencp=cmcsc10 scaled\magstep3

\newif\ifCONF \CONFfalse
\newif\ifBREAK \BREAKfalse
\newif\ifsectionskip \sectionskiptrue

%
%%%%%%%%%%%%%%%%%%%%%%%%%%%%%%%%%%%%%%%%%%%%%%%%%%%%%%%%%%%%%%%%%%%%%%
%%
%           FORMATS
%%%%%%%%%%%%%%%%%%%%%%%%%%%%%%%%%%%%%%%%%%%%%%%%%%%%%%%%%%%%%%%%%%%%%%
%%
%
% Nuclear Physics Proceedings Format
%
%
\def\NuclPhysProc{
\let\lr=L
\hstitle=8truein\hsbody=4.75truein\fullhsize=21.5truecm\hsize=\hsbody
\hstitle=8truein\hsbody=4.75truein\fullhsize=20.7truecm\hsize=\hsbody
\output={
  \almostshipout{\leftline{\vbox{\makeheadline
  \pagebody\makefootline}}}\advancepageno
     }
\def\papersize{\SIZE\OFFSET\skip\footins=\bigskipamount}
\def\SIZE{\hsize=10.0truecm\vsize=27.0truecm}
\def\OFFSET{\voffset=-1.4truecm\hoffset=-2.40truecm}
\message{NUCLEAR PHYSICS PROCEEDINGS FORMAT}
\def\makeheadline{
\iffrontpage\line{\the\headline}
             \else\vskip .0truecm\line{\the\headline}\vskip .5truecm
\fi}
\def\makefootline{\iffrontpage\vskip  0.truecm\line{\the\footline}
               \vskip -.15truecm\line{\the\date\hfil}
              \else\line{\the\footline}\fi}
\paperheadline={\hfil}
\paperstyle}
%
%
% World Scientific Format
%
%

%
%
% Johns Hopkins
%
%

%
%
% Reprint Volume Format
%
%
\def\ReprintVolume{\smallsize
\def\papersize{\hsize=18.0truecm\vsize=23.1truecm\voffset -.73truecm
    \hoffset -.65truecm\skip\footins=\bigskipamount
    \normaldisplayskip= 20pt plus 5pt minus 10pt}
\message{REPRINT VOLUME FORMAT}
\paperstyle\baselineskip=.425truecm\parskip=0truecm
\def\makeheadline{
\iffrontpage\line{\the\headline}
             \else\vskip .0truecm\line{\the\headline}\vskip .5truecm
\fi}
\def\makefootline{\iffrontpage\vskip  0.truecm\line{\the\footline}
               \vskip -.15truecm\line{\the\date\hfil}
              \else\line{\the\footline}\fi}
\paperheadline={
\iffrontpage\hfil
               \else
               \tenrm\hss $-$\ \folio\ $-$\hss\fi    }
\def\sectionfont{\bf}    }
%
%
% CERN-preprint format (default)
%
%
\def\SIZE{\hsize=15.73truecm\vsize=23.11truecm}
\def\OFFSET{\voffset=0.0truecm\hoffset=0.truecm}
\message{DEFAULT FORMAT}
\def\papersize{\SIZE\OFFSET\skip\footins=\bigskipamount
\normaldisplayskip= 35pt plus 3pt minus 7pt}
\Pubnum={\rm \the\pubnum }
\def\title#1{\vskip\frontpageskip\vskip .50truein
     \titlestyle{\seventeencp #1} \vskip\headskip\vskip\frontpageskip
     \vskip .2truein}
\def\author#1{\vskip .27truein\titlestyle{#1}\nobreak}
\def\andauthor{\vskip .27truein\centerline{and}\author}
\def\p@bblock{\begingroup \tabskip=\hsize minus \hsize
   \baselineskip=1.5\ht\strutbox \topspace+2\baselineskip
   \halign to\hsize{\strut ##\hfil\tabskip=0pt\crcr
  \the \Pubnum\cr}\endgroup}
\def\makefootline{\iffrontpage\vskip .27truein\line{\the\footline}
                 \vskip -.1truein
%\line{\the\date\hfil}
              \else\line{\the\footline}\fi}
\paperfootline={\iffrontpage\message{FOOTLINE}
% \the\Pubnum
\hfil\else\hfil\fi}

\def\abstract{\vskip2\frontpageskip\centerline{\twelvebf Abstract}
              \vskip\headskip }

\paperheadline={
\iffrontpage\hfil
               \else
               \twelverm\hss $-$\ \folio\ $-$\hss\fi}
%
% Journal abbreviations
%
\def\nup#1({\refbreak\ Nucl.\ Phys.\ $\underline {B#1}$\ (}
\def\plt#1({\refbreak\ Phys.\ Lett.\ $\underline  {#1}$\ (}
\def\cmp#1({\refbreak\ Commun.\ Math.\ Phys.\ $\underline  {#1}$\ (}
\def\prp#1({\refbreak\ Physics\ Reports\ $\underline  {#1}$\ (}
\def\prl#1({\refbreak\ Phys.\ Rev.\ Lett.\ $\underline  {#1}$\ (}
\def\prv#1({\refbreak\ Phys.\ Rev. $\underline  {D#1}$\ (}
\def\und#1({            $\underline  {#1}$\ (}
%
% Line breaks
%

\def\rB{\hfil\penalty1000\hfilneg}
%
% Hyphenations
%
\hyphenation{sym-met-ric anti-sym-me-tric re-pa-ra-me-tri-za-tion
Lo-rentz-ian a-no-ma-ly di-men-sio-nal two-di-men-sio-nal}
%
%
% Miscellaneous macros
%
%

\def\coeff#1#2{{\textstyle { #1 \over #2}}\displaystyle}
\def\boxit#1{\vbox{\hrule\hbox{\vrule\kern3pt
\vbox{\kern3pt#1\kern3pt}\kern3pt\vrule}\hrule}}
\message{ by V.K, W.L and A.S}
\catcode`@=12
\paperstyle

\def\chi {\X}

\def\pie{\rrr\pie{
D.M.~Pierce, {\it  A (1,2) Heterotic String with Gauge Symmetry.}
\prv53 (1996) 7197.}}
\def\FMS {\rrr\FMS {D.~Friedan, E.~Martinec and S.~Shenker,
{\it Conformal Invariance, Supersymmetry and String Theory.}
 \nup271 (1986) 93.}}
\def\LeL{\rrr\LeL{
W.~Lerche and D.~L\"ust,
{\it Covariant Heterotic Strings and Odd Selfdual Lattices.} \plt  B187 (1987)
45.}}
\def\KLLS{\rrr\KLLS{A.~Kostelecky, O.~Lechtenfeld, W.~Lerche,
 S.~Samuel and S.~Watamura, {\it Conformal Techniques, Bosonization and Tree
Level String Amplitudes.}
\nup238 (1987) 173.}}
\def\LLSB{\rrr\LLSB{
W.~Lerche, D.~L\"ust and A.N.~Schellekens,\rB
{\it Chiral Four-dimensional Heterotic Strings from
Self-dual Lattices.}
\nup287 (1987) 477.}}
\def\LSWP{\rrr\LSWP{W.~Lerche, A.~Schellekens and N.~Warner,
{\it Lattices and Strings},
\prp 177 (1989) 1.}}
\def\oov{\rrr\oov{
H. Ooguri and C. Vafa, {\it N=2 Heterotic Strings.} \nup367 (1991) 83.}}
\def\oova{\rrr\oova{
H. Ooguri and C. Vafa, {\it Geometry of N=2 Strings.} \nup361 (1991) 469}}
\def\KMa{\rrr\KMa{
D. Kutasov and E. Martinec, {\it New principles for string/membrane
unification.} \nup447 (1996) 652.}}
\def\KMb{\rrr\KMb{
D. Kutasov, E. Martinec and M. O'Loughlin, {\it Vacua of M-theory and N=2
strings.} \nup477 (1996) 675.}}
\def\KMc{\rrr\KMc{
D. Kutasov, E. Martinec, {\it M-Branes and N=2 Strings.}, hep-th/9612102.}}
\def\green{\rrr\green{
M. Green, {\it World sheets for world sheets.}, \nup B293 (1987) 593.}}
\def\BGR{\rrr\BGR{
L. Baulieu, M. Green and E. Rabinovici, 
{\it Superstrings from theories with $N>1$ world-sheet supersymmetry.}
to be published in Nuclear Physics B .}}
\def\SchH{\rrr\schH{A.N. Schellekens,
{\it Classification of Ten-dimensional Heterotic Strings.}
\plt B277 (1992) 277.}}
\def\SchM{\rrr\SchM{A.N.~Schellekens,
{\it Meromorphic c=24
Conformal Field Theories},
\cmp 153 (1993)  159.}}
\def\Mont{\rrr\Mont{P.~Montague,
{\it Orbifold constructions and the classification of selfdual c = 24 conformal
field theories},
\nup 428 (1994) 233.}}
\def\DGM {\rrr\DGM  {L.~Dolan, P.~Goddard and P.~Montague,
{\it Conformal field theory, Triality and the Monster group}
\plt B236 (1990) 165.}}
\def\CFQS{\rrr\CFQS{A.~Cohn, D.~Friedan, Z.~Qiu and S.~Shenker,
{\it Covariant quantization of supersymmetric string theories: the spinor field
of the Ramond-Neveu-Schwarz model.}
\nup278 (1986) 577.}}

\paperstyle

\def\half{\coeff12}

\def\Zbf{{\bf Z}}

\def\X{{\cal X}}

\catcode`@=11
\def\ninef@nts{\relax
    \textfont0=\ninerm          \scriptfont0=\sixrm
      \scriptscriptfont0=\sixrm
    \textfont1=\ninei           \scriptfont1=\sixi
      \scriptscriptfont1=\sixi
    \textfont2=\ninesy          \scriptfont2=\sixsy
      \scriptscriptfont2=\sixsy
    \textfont3=\tenex          \scriptfont3=\tenex
      \scriptscriptfont3=\tenex
    \textfont\itfam=\nineit     \scriptfont\itfam=\seveni  % no
\sevenit
    \textfont\slfam=\ninesl     \scriptfont\slfam=\sixrm % no
\sevensl
    \textfont\bffam=\ninebf     \scriptfont\bffam=\sixbf
      \scriptscriptfont\bffam=\sixbf
    \textfont\ttfam=\tentt
    \textfont\cpfam=\tencp }
\def\ninepoint{\ninef@nts \samef@nt \b@gheight=9pt \setstr@t }
\newif\ifnin@  \nin@false
\def\Tenpoint{\tenpoint\twelv@false\nin@false\spaces@t}
\def\Twelvepoint{\twelvepoint\twelv@true\nin@false\spaces@t}
\def\Ninepoint{\ninepoint\twelv@false\nin@true\spaces@t}
\def\spaces@t{\rel@x
      \iftwelv@ \ifsingl@\subspaces@t3:4;\else\subspaces@t1:1;\fi
       \else \ifsingl@\subspaces@t3:5;\else\subspaces@t4:5;\fi \fi
      \ifdoubl@ \multiply\baselineskip by 5
         \divide\baselineskip by 4 \fi
       \ifnin@ \ifsingl@\subspaces@t3:8;\else\subspaces@t4:7;\fi \fi
}
\def\Vfootnote#1{\insert\footins\bgroup
   \interlinepenalty=\interfootnotelinepenalty \floatingpenalty=20000
   \singl@true\doubl@false \iftwelv@ \Tenpoint
   \else \Ninepoint \fi
   \splittopskip=\ht\strutbox \boxmaxdepth=\dp\strutbox
   \leftskip=\footindent \rightskip=\z@skip
   \parindent=0.5\footindent \parfillskip=0pt plus 1fil
   \spaceskip=\z@skip \xspaceskip=\z@skip \footnotespecial
   \Textindent{#1}\footstrut\futurelet\next\fo@t}

\def\small#1{\vskip .3truecm\footnoterule\nobreak
\Ninepoint\parindent=2pc\sl
\hang #1 \vskip .3truecm\nobreak\footnoterule}

\def\small#1{}

%%%%%%%%%%%%%%%%%%%%%%%%%%%%%%%%%%%%%%%%%%%%%%%%%%%%%%%%%%%%%%%%%%%%%%
  
%%%%%%%%%%%%%%%%%%%%%%%%%%%%%%%%%%%%%%%%%%%%%%%%%%%%%%%%%%%%%%%%%%%%%%

\pubnum={{}}
\rightline{NIKHEF 97-023}
%\rightline{hep-th/96010..}
\rightline{hep-th/9706122}
\rightline{March 1997}
\date{June 1997}
\pubtype{CRAP}
\titlepage
\message{TITLE}

\title{\fourteenbf On the classification of (2,1) heterotic strings }
\author{E. Rabinovici\foot{eliezer@vms.huji.ac.il. Partialy supported
by the American-Israeli
BiNational Science foundation(BSF) and the center of excellency  sponsored
by the Israeli Science Fund.}}
\line{\hfil
 {\it Racah Institute of Physics Hebrew University, Jerusalem,
Israel}
 \hfil}
\andauthor{A. N. Schellekens\foot{t58@attila.nikhef.nl}}
\line{\hfil \it NIKHEF-H, P.O. Box 41882, 1009 DB Amsterdam,
The Netherlands  \hfil}
\bigskip

\abstract \noindent
We classify all untwisted (2,1) heterotic strings. The only solutions
are the three already known cases, having massless spectra consisting
either of 24 chiral fermions, or of
24 bosons, or of 8 scalars and 8 fermions of each chirality.

   \baselineskip= 15.0pt plus .2pt minus .1pt

\chapter{Introduction}

Closed string theories with N=0 or N=1 world-sheet supersymmetries
can be completely
classified in their maximal (``critical") dimension, assuming only
(super)con\-for\-mal and modular invariance. This classification yields
a single (0,0) string, four (1,1) strings (IIA, IIB, plus two theories
without space-time supersymmetry), and nine (0,1) strings (the two
heterotic superstrings, plus seven non-supersymmetric theories). In
this paper we want to extend these results to (a subclass of) N=2
heterotic strings.

The basic idea behind such classifications is that in the maximal
dimension at least one of the chiral sectors of the underlying
two-dimensional conformal field theory has its central charge saturated
by fields with space-time indices, leaving no room for an unknown
internal conformal field theory. The only non-trivial structure one
can have in this saturated sector are then the spin-structures of the
world-sheet fermions, which have known modular transformations among
themselves.
In such a situation it is always possible to map
the modular invariant partition to a meromorphic one (\ie\ to a
character of a rational conformal field theory with a single primary
field; such theories have $c=8k$).
If the central
charge of this meromorphic conformal field theory is 8, 16 or 24,
one can use the available classifications of meromorphic conformal
field theories to determine all possibilities for the unknown chiral
internal sector. The (1,1) strings map to $c=16$ theories, whereas
the (0,1) strings map to $c=24$ theories. In the former case there are
no internal sectors, but there are several ways of combining the
spin structures, which are easily read off from the two meromorphic
$c=16$ theories, $E_8 \times E_8$ and $Spin(32)/\Zbf_2$. In the latter
case all possibilities for the internal $c=16$ conformal field theory
as well as all ways of combining it with the NSR spin-structures can be
read off \SchH\ from the list of possible meromorphic $c=24$ conformal
field theories \SchM.

Similar considerations should apply to strings with N=2 world-sheet
supersymmetry. Indeed, Ooguri and Vafa \oov\ gave a classification
for N=2 strings under certain assumptions. However they made
the unnecessarily restrictive
assumption that the internal conformal field theories are
essentially torus compactifications. Modular invariance then only
allows the 24 Niemeier lattices for the right sector of (2,0) strings
and the
$E_8$ torus for (2,1) strings.
Later \multref\pie{\KMa\KMb\KMc}\ more general solutions were found in the
latter type of
theories, but
without claims to completeness. The classification of (2,0) strings
is easily completed by replacing the 24 Niemeier lattices by any
of the 71 meromorphic conformal field theories enumerated in \SchM\
(many of which have been explicitly constructed, see \eg\ \doubref\DGM\Mont).

In  \KMa,\KMb\ and \KMc\
target spaces of N=2 strings were constructed. A large variety of
such theories was uncovered. The identification of the exact nature
of these target spaces was aided by a conjecture that they provide
world sheet theories of critical strings (as well as world volume
theories for 3-d membranes). This realized the
idea proposed in \green. It
is of interest to arrive at a complete classification
of these theories.

There are two additional features in N=2 heterotic strings that can
complicate the analysis: the moduli integration in
the left (N=2) sector, which identifies the spin structures, and the
need for a ``null current" reduction of an extra space and time direction
in the right sector. When fully exploited, these features may give
rise to additional possibilities of combining left- and rightmovers
($Z_n$-strings \oova) or twisted boundary conditions in the null
directions \KMa. Here we will consider only the simplest case, where
such additional twists are absent. Perhaps our analysis can be
generalized to cover the other cases as well.

Under the assumptions stated above the left sector is modular invariant
by itself, and requires no further discussion.
The null current reduction requires only slightly more care.
A (2,2) string has a (real) critical dimension equal to four, and
the no-ghost theorem requires the space-time metric signature to be
$(+,+,-,-)$. To build heterotic strings,
one chiral sector of such a theory has to be combined with
N=0 or N=1 chiral sectors.
The presence of
two time directions in the N=2 string target space requires the
same in the N=0 or N=1 sectors. Furthermore
the presence of a $U(1)$ gauge symmetry in the N=2 sector requires
the same in the N=0 or N=1 sectors \oov\  
%, since vector fieldscannot be chiral in two dimensions
(in the case of N=1, supersymmetry requires  in addition to the
vector current also a fermionic current in that sector).
Ooguri and Vafa showed
how these two changes essentially cancel each other:
the extra symmetries lead to extra ghosts, whose contribution to the
conformal anomaly
leads to an increase of the critical dimension by two ($d=28$ and $d=12$
for N=0 and N=1 respectively), the no-ghost theorem requires one of
these extra dimensions to be time-like, making a combination with a left
N=2 string possible, and finally BRST invariance requires the $U(1)$ current
to be a ``null current". This means that these currents are of the
form $\nu_{\mu} \partial X^{\mu}$  (plus $\nu_{\mu} \psi^{\mu}$ for
N=1 theories) where $\nu$ is a light-like vector. The gauge symmetry
then implies that physical excitations must have momenta in a plane
orthogonal to the null vector $\nu_{\mu}$, and that all momenta that
differ by $\nu_{\mu}$ are identified. In this way one recovers 26 and
10-dimensional Lorentz invariance for (0,0) and (1,1) strings respectively.
The (2,1) and (2,0) strings have two or three-dimensional Lorentz
invariance, depending on the orientation of the vector $\nu^{\mu}$.
We will only consider two dimensions here. For a discussion of the
three dimensional case, given a
modular invariant two-dimensional theory, we refer to Appendix A of \KMc.

The assumption that there are no twists in the null directions
implies that, apart from the null-current constraint, the theories
we consider are Lorentz-invariant in four dimensions. This constrains
four bosons of the $c=28$ N=0 sector and four bosons and four fermions
of the $c=18$ N=1 sector, leaving respectively a c=24 conformal field
and a c=12 superconformal field theory undetermined. It is these theories
we wish to classify.

One can see that the c=24 N=0 theory must be modular
invariant by itself, and a classification of such theories is already
available, as mentioned above. The c=12 superconformal field theory
makes a contribution to the partition function that depends on
the spin structure of the right-moving world-sheet fermions. It must
thus be a theory with four characters with modular transformations
dictated by those of the NSR fermions.

To classify the (2,1) strings we will make use of techniques
similar to those used in \SchH\ for the classification of ten-dimensional
heterotic strings. The starting point is a bosonic formulation of
all fermions with space-time indices and all bosonic ghosts, \ie\ all
fields carrying non-trivial spin-structures. We emphasize that
nothing is assumed about the other right-moving degrees of freedom,
except that they should form a $c=12$ superconformal field theory.
We will in fact weaken this requirement to ``conformal", and inspect
superconformal invariance at the end.

For a ten-dimensional
heterotic string the bosonization of $\psi^{\mu}$ and
the $\beta,\gamma$ ghosts yields a description in terms of a
six-dimensional
``covariant
lattice" with a metric with signature $(+,+,+,+,+,-)$ \CFQS\
(see also \multref\KLLS{\LeL\LSWP}). The five
positive metric fields correspond to the bosonized $SO(9,1)$ NSR
fermions while the last component corresponds to
the bosonized superghosts (note that the lattice
metric is not related to the space-time metric). If one ignores the
metric, the lattice is just the weight lattice of $D_6$, but to indicate
its metric we will call it $D_{5,1}$.
The lattice structure
is a direct consequence of the requirement that all 10 space-time
fermions as well as the superghosts must have the same spin structures
on any Riemann surface. The generalization to lower dimensions is
straightforward, and in two dimensions one needs a two-dimensional
lattice $D_{1,1}$ with signature $(+,-)$.

In the present case some changes are required, due to the null current
and the extra ghosts. A useful guiding principle is the fact that such
a formulation is also available for (0,1) strings, i.e. the usual
heterotic strings, and should give the same answer. This alternative
formulation is in (28,12) dimensions rather than (26,10), and
includes null currents and extra ghosts.
If one takes the equivalence between these formulations of (0,1)
strings
for granted the main
result for (2,1) strings follows very easily.
However we prefer not to take it for granted,
and examine the new covariant lattice description more carefully.
We will present this in such a way that
the results are valid
for a (compactified) (0,1) string in $2n$ flat space-time
dimensions as well as a (2,1) string in 2 flat space-time dimensions
(for $n=1$).

The right-moving sector of such a string theory is built out of
a $c=12$ superconformal field theory, multiplets of bosons and fermions
$X^{\mu}, \psi^{\mu}$, where $\mu$ is a space-time index of a space
with signature ($2n,2$), plus ghosts. The ghost sector consists of
reparametrization ghosts $b$ and $c$, superghosts $\beta$ and $\gamma$,
plus fermionic ghosts $b'$ and $c'$ of the
world-sheet $U(1)$ gauge symmetry and their
bosonic superpartners $\beta'$ and $\gamma'$.
The following table summarizes the dimensions and total central charge
of these fields.
\thicksize=0pt
\vskip .7 truecm
\begintable
fields | dimensions | central charge \cr
$(b,c)$ | $(2,-1)$  | -26 \nr
$(\beta,\gamma)$ | $(\coeff32,-\coeff12)$  | 11 \nr
$(b',c')$ | $(1,0)$  | -2 \nr
$(\beta',\gamma')$ | $(\coeff12,\coeff12)$  | -1
\endtable
\vskip .7 truecm
In comparison with the right sector of the
usual (0,1) string there are two extra
world-sheet fermions plus an extra bosonic ghost system $(\beta',\gamma')$
that depend on the spin structure. World-sheet supersymmetry requires
all these spin structures to be identical. Hence if we bosonize the
fermions and the bosonic ghosts, the resulting two additional
bosonic degrees of freedom can be taken into account by adding
two dimensions to the covariant lattice, with opposite metric. So the
right-moving fermionic sector is described by a covariant
lattice $D_{n+1,2}$ plus the internal $c=12$ conformal field theory.

In the one-loop partition functions
the fermionic ghosts $b,c,b',c'$ cancel the contribution of four bosons
$X^{\mu}$, reducing their contribution to the transverse one, which in
two dimensions is trivial.
The superghosts $(\beta,\gamma)$ cancel the contributions of two
fermions $\psi^\mu$ of opposite signature, exactly as in the (0,1)
string, while the remaining bosonic ghosts cancel the
contribution of the remaining two fermions $\psi^\mu$. At arbitrary
genus, the partition function has the form
$$ \sum_{\alpha} (Det_{1/2}^{\alpha})^{2n} (Det_{1/2}^{\alpha})^{-1}
 (Det_{3/2}^{\alpha})^{-1} {\cal C}_{\alpha} \eqn\detcomb $$
where $\alpha$ denotes the spin structure, and $\cal C$ the
contribution of the unknown internal CFT. Note that the extra fermion
and the extra ghost just cancel, since both are $(\half, \half)$
determinants, albeit with opposite metrics.
This cancellation is exact at arbitrary genus, unlike the cancellation
of the spin-$\coeff32$ determinant, which occurs only for genus 1.
The exact cancellation of the $\beta'\gamma'$ ghost contribution is
essential for the exact equivalence of the two (0,1) formulations discussed
above.
Consequently the discussion
of modular invariance is unchanged. Note also that the
conjugacy classes on the $D_{n+1,2}$ lattice have the same norms (modulo
even integers) as those of the $D_{n,1}$ lattice used in
the standard formulation of two-dimensional
(0,1) strings.

Consider now $n=1$, either for (2,1) strings, or for (0,1) strings
with a meromorphic CFT as their left-moving sector.
Since the left sector
is separately modular invariant, the combination \detcomb\ with $n=1$
must be modular invariant as well.
It is known (see \eg\ \LSWP) that the ratios
$$ Y^{\alpha}={Det_{1/2}^{\alpha} \over
 Det_{3/2}^{\alpha}} \eqn\Ya $$
transform in exactly the same way as
$$ X^{\alpha}=(Det_{1/2}^{\alpha})^4 \sum_{\beta} (Det_{1/2}^{\beta})^8
\eqn\Xa $$
Hence the following combination must be modular invariant
$$ \sum_{\alpha}(Det_{1/2}^{\alpha})^{4}{\cal C}^{\alpha}\sum_{\beta}
(Det_{1/2}^{\beta})^8\ . $$
This combination can be interpreted as the partition function of a
meromorphic $c=24$ theory, since only determinants of ordinary fermions
occur, and since ${\cal C}$ represents a unitary CFT.

The fermion determinants can in fact be interpreted as characters
of the level-1 affine algebras $D_4$ and $E_8$. To read off the
possibilities for the internal CFT one must look for meromorphic
$c=24$ CFT's that have $D_4 \times E_8$ at level 1 as a subalgebra.
Due to the presence of the $E_8$ factor this problem reduces to looking
for $D_4$ subalgebras of $c=16$ meromorphic CFT's, and then the
only possibilities are the familiar even
self-dual lattices $E_8 \times E_8$ and $D_{16}$.

To obtain level 1, $D_4$ must be embedded in just one $E_8$ factor, and
then there is just one possibility, namely the embedding
$E_8 \supset D_4^{\rm ghost} \times D_4^{\rm int}$ defined by
$$ (248) \to (28,1)+(1,28)+(8_v,8_v)+(8_s,8_s)+(8_c,8_c) \eqn\Dec$$
On the other hand, for the embedding
$D_4^{\rm ghost} D_{12}^{\rm int}\subset D_{16}$ there are
three distinct possibilities, namely
$$ (496) \to (28,1)+(1,276)+(8_v,24) \eqn\Deca$$
$$ (496) \to (28,1)+(1,276)+(8_s,24) \eqn\Decb$$
$$ (496) \to (28,1)+(1,276)+(8_c,24) \eqn\Decc$$
These three embeddings are related by triality. One might think
that the same possibility exists also in the first case, but there
all triality rotated embeddings are in fact indistinguishable, because
they can be undone by a compensating triality rotation in the second $D_4$
factor.

The spectrum is now easy to obtain using the rules formulated in \LLSB.
These rules require some more discussion due to the extra components
on the lattice. Again the (28,12) dimensional formulation of (0,1)
strings can serve as a guiding principle.
Considering again first (0,1) strings
in arbitrary (even) dimensions $D=2n$. The two formulations
involve then covariant lattices $D_{n,1}$ and $D_{n+1,2}$
respectively.

Let us first review the argument for lattices $D_{n,1}$.
In terms of lattices, the partition function map
replacing $Y_{\alpha}$ by $X_{\alpha}$ (\cf\ \Xa, \Ya) is
equivalent to replacing $D_{n,1}$ or $D_{n+1,2}$ by $D_{n+3} \times E_8$, using
a map on conjugacy classes rather then on individual vectors.
Not all
vectors on the $D_{n,1}$ lattice correspond to physical states, since
the ghost charge can be changed by acting with the picture changing
operator $e^{i\phi} T_F$. Here $\phi$ is the boson in terms of which
$\beta$ and $\gamma$ are bosonized, and $T_F$ the supercurrent.
The simplest picture is the ``canonical" one, in which the physical
states are in one-to-one correspondence with the negative modes of the
bosonic and fermionic oscillators acting on the ground state.
To make sure that the positive modes of the bosonic ghosts
$\beta$ and $\gamma$
annihilate the ground state one must assign a ghost charge $q$ to it;
otherwise the action of these modes would render the energy unbounded from
below. As discussed in \FMS, this charge is
$$\eqalign{ q &=\half-\lambda ~~~~~~~~~~~~~~~ (\hbox{NS})\cr
            q &=1-\lambda~~~~~~~~~~~~~~~(\hbox{R}) \ .\cr}\eqn\VacCharge $$
for a general
$\beta, \gamma$ ghost system of conformal weights $\lambda, 1-\lambda$.
For the superghost system this yields $q=-1$ (NS) and $q=-1/2$ (R).
To read off physical
states one only considers vectors on $D_{n,1}$ whose last components are
equal to one of these values. To read off the light cone states directly
one may furthermore fix the second-to-last entry to 0 (NS) or $-\half$ (R),
so that one only considers vectors with last components equal to
$(0,-1)$ or $(-\half,-\half)$. All those vectors are precisely obtained
by considering the complements of the conjugacy classes $(v)$ and $(s)$ of
$D_4 \subset D_{n+3}$.

Similar considerations apply
for the $D_{n+1,2}$ lattice, except that now there is
a second picture changing operator associated with the $\beta',\gamma'$
bosonic ghost system.
In comparison with the lattice of the light cone rotation group $D_{n-1}$
the covariant lattice $D_{n+1,2}$ has four extra components, corresponding
respectively to
\item\dash (a) The extension from light cone rotation group $SO(2n-2)$ to
 the Lorentz group $SO(2n-1,1)$.
\item\dash (b) The null current direction.
\item\dash (c) The bosonized $\beta, \gamma$ ghosts.
\item\dash (d) The bosonized $\beta', \gamma'$ ghosts.

\noindent
Entries (a) and (c) are exactly as before.

The $\beta', \gamma'$ ghosts can be treated completely analogously to
the $\beta, \gamma$ ghosts. Setting $\lambda=\half$ in \VacCharge\ we find
now $q'=0$ (NS) and $q'=\half$ (R). This determines entry (d). Entry (b)
is fixed by the null-current constraint on the vertex operator. For
the NS and R massless states this fixes entry (b) to (0) and ($\half$)
respectively, \ie\ the projection on the null current direction should
respectively be a singlet or a spinor of definite chirality (the choice
of chirality is irrelevant). We have verified that these ghost charge
assignments and constraints are in agreement with BRST invariance of
the vertex operator. They are also in agreement with \KMb.

Combining all this, we find thus that
lightcone degrees of freedom can be read off by fixing the last four
components to either $(0,0,-1,0)$ for NS or $(-\half,+\half,-\half,+\half)$
for R. After mapping to $D_{n+3}$ this corresponds precisely to the
conjugacy classes $(v)$ and $(s)$ of $D_4$, exactly as in the other
formulation in terms of $D_{n,1}$ covariant lattices.

Since we are considering two-dimensional target spaces, using light cone
states is not as convenient as in higher dimensions.
However, it follows from the foregoing
discussion that one may also read off the covariant states directly
by stripping off the conjugacy classes $(v)$ and $(s)$ of $D_3$ (rather than
$D_4$). The massless spectra obtained from \Dec -- \Decc\ are then
respectively 8 (non-chiral) scalars + 8 chiral fermions + 8 anti-chiral
fermions; 24 (non-chiral) scalars; 24 chiral fermions or 24 anti-fermions.
These particles can be assigned to representations of the internal group,
which however is not dynamically realized. For \Dec\ any permutation of
the three non-zero conjugacy classes of $SO(8)$ is a possible assignment,
whereas of course for \Deca -- \Decc\ all states are in the vector
representation of $D_{12}$. The correct counting of the fermionic
states also follows from table 6 in \LSWP, where the number of
gravitini in (0,1) strings is displayed. In the case of (2,1) strings these
fermions lose their space-time index $\mu$ and become spin-{$\coeff12$}
fermions.

So far we have not used the requirement of superconformal invariance,
but only modular invariance and conformal invariance. If we had found
new solutions, we should check that they are consistent with
world sheet supersymmetry. However,
these three cases are not new, and
have all been discussed before in \pie\ and \KMa,\KMb,\KMc.
In these papers these solutions were obtained {\it assuming} a
free fermionic description of the $c=12$ conformal field theory. What
we have shown is that this assumption is unnecessary, and that no
other solutions exist, independent of a particular construction method.

The identification of the explicit target space lagrangian giving rise to the
physical spectra we have found in three cases is more involved.
The (2,0) world sheet theories were conjectured (in \KMa) to lead to purely
bosonic compatifications in target space. The system containing 24 physical
seems  (\KMc)
to represent a special (1,1) string theory. The other
two spectra, the 8 bosons and eight fermions and the 24 chiral(or anti-chiral)
fermions were associated with a type IIB string and a (2,1) string
respectivly. The fact that a (2,1) system reproduces itself suggests
some flow in the space of self-reproducing theories. Other work (\BGR)
suggests that actualy all $(N,M)$ systems with $N,M<5$ should be reproduced
as target space descriptions of (2,1) worldsheet theories. In \KMc\ it is
claimed that (1,0) and (2,2) theories can be identified as well. The former
 is found using an orbifold twist in the null direction,
 a possibility which we
have not considered, and the latter is obtained from a partition
function that violates the spin-statistics relation. The
$(2,2)$ target space contains no physical states and is in fact
a topological theory, so perhaps this is just what is needed,
with the wrong-statistics fields interpreted as ghosts. However,
in any case our method can not produce such a solution, since it is based
on a map to a bosonic string partition function with positive signs, and
since furthermore
very little is known about modular invariants with non-definite signs.
It also not at all clear how to the data
on only the physical states will differentiate between (2,2),(3,3) or (4,4)
topological systems. We wish to return to this problem in the future.

\ack
E.R. is indebted to the physics department
of the university of Utrecht, where
most of this work was done, for its hospitality.

\par \penalty-4000\vskip\chapterskip
   \spacecheck\referenceminspace \immediate\closeout\referencewrite
   \referenceopenfalse
   \line{\fourteenrm\hfil REFERENCES\hfil}\vskip\headskip
   \endlinechar=-1
   \input referenc.texauxil
   \endlinechar=13
   
\end